# Supernova observations fit Einstein-deSitter expansion in 4-sphere

Tuomo Suntola
tuomo.suntola@sci.fi

and

Robert Day
bobday.nh@verizon.net

ABSTRACT

Recent observations on the redshift and magnitude of distant supernovae have raised fundamental questions of the cosmological structure and the development of space. Interpretation of the observations with the standard cosmology model suggests accelerating expansion of space which requires an assumption of unknown dark energy working against gravitation at cosmological distances. This paper presents an analysis which shows that the observed magnitude redshift relationship is in an excellent agreement with the prediction derived by assuming a zero energy balance of motion and gravitation in an expanding 4-sphere. Such an approach can be regarded as Einstein-deSitter expansion in the fourth dimension, suggesting a decelerating expansion velocity until zero in infinity.

Subject headings: large-scale structure of universe — cosmological parameters — galaxies: distances and redshifts — cosmology: theory — cosmology: observations

## 1. Introduction

Over the past several years, studies, such as those of Perlmutter[Ref. 1], Knop[2], and Riess[3] of the distance modulus vs. redshift of Type Ia supernovae have presented models, based on the standard theory of cosmology, that require an acceleration in the expansion rate of the universe in order to adequately fit the data. The idea that the universe is expanding at an accelerating rate has now been accepted by most cosmologists, even though there is little real detectable evidence for an underlying mechanism, such as dark energy, that might drive the acceleration. Here we show that the equation

$$m = m_0 + 5\log_{10} z + 2.5\log_{10}(z+1) \qquad (1)$$

where $m$ is the apparent magnitude and $m_0$ a reference magnitude, provides an excellent model of the existing SN Ia magnitude vs. redshift data. This equation can be derived by assuming a zero energy balance of motion and gravitation in space spherically closed through the fourth dimension. Such a solution does not resort to an assumption of acceleration of cosmic expansion nor to assumptions on $\Omega_m$, $\Omega_\lambda$, or the Hubble constant $H_0$ [4,5]. Spherically closed dynamic space converts spacetime with variable time and distance coordinates into universal coordinates by adding the effect of the motion of space in the fourth dimension. The motion of space also eliminates the need of a cosmological constant, originally proposed by Einstein to prevent the collapse of spherically closed static space [6].

## 2. Data Used for this Study

The data used for this study is that reported in Table 5 of Riess et al's recent paper "Type Ia Supernova Discoveries at z > 1 From the Hubble Space Telescope: Evidence for Past Deceleration and Constraints on Dark Energy Evolution", The Astrophysical Journal, 2004, 607, 665-687. The measurement for SN 1991ag was omitted from this paper, but was reported in a preprint version. This measurement was judged to have been inadvertently omitted and was included for this study. Table 5 in Riess et al's paper includes measurements not only for recent SNe Ia discoveries, but also the measurements they deemed reliable from previous SNe Ia studies (see the discussion in Section 3.1 of their paper). Riess et al divide the data into two classes, a "gold" dataset consisting of 157 objects which they label as "high-confidence" SNe Ia, and a "silver" dataset consisting of 29 objects to which they give the label "likely but not certain" SNe Ia.

### 3. Procedure

### 3.1 Data Verification

First, we manually checked our transcription of Reiss et al's data, and, as an additional check, we verified that we could reproduce their results. We used equations 11 and 12 in Riess et al to perform our calculations:

$$d_L = \frac{c(1+z)}{H_0} \int_0^z \frac{1}{\sqrt{(1+z)^2(1+\Omega_m z) - z(2+z)\Omega_\lambda}} dz \qquad (2)$$

and

$$\mu_p = m - M = 5 \log_{10} d_L + 25 \qquad (3)$$

where $\mu_p$ is referred to as the extinction-corrected distance modulus.

Equations (2) and (3) can be derived from the standard cosmology model, and they will be referred to as the "standard cosmology equations" in this study. In an email communication, Dr. Riess stated that to produce the Hubble diagram in Figure 4 of their paper (which included the gold dataset plus the two silver HST objects), he used the parameter values $\Omega_m = 0.29$, $\Omega_\lambda = 0.71$, and $H_0 = 64.5$. For these values, we compute a $\chi^2$ of 177.4117, which is consistent with results Riess et al report.

In Table 4 of their paper, Riess et al report obtaining a "$\chi^2$ (for 157 SNe Ia)" of 178 when $\Omega_m$ is set to 0.27 and $\Omega_\lambda$ is set to 0.73. Presumably, they are referring to the gold dataset in Table 4, since the gold dataset contains 157 points. We compute an optimum $\chi^2$ of 178.1727 for those omega values when $H_0$ is set to 64.9. Since the $\chi^2$ values we obtained are consistent with those in Riess et al, we deemed our data correct. However, we noted the low value needed for $H_0$, and will comment further on that later in this study.

### 3.2 Comparing the Fits

Next, we compared the fits of Equation 1 and standard cosmology equations to the data. Figure 1 shows the optimum fits of Equation 1 and the standard cosmology equations to the data Riess et al used for their Figure 4. The fit for Equation 1 is represented by the blue line and that for standard cosmology equations is the red line, slightly above. The optimum $\chi^2$ for standard cosmology equations is 177.1079, at $\Omega_m = 0.31$, $\Omega_\lambda = 0.69$, and $H_0 = 64.3$. The optimum $\chi^2$ for Equation 1 is 197.7351, obtained for $\mu_p = 43.43$.

Figure 2 shows how the optimum fit of Equation 1 to the data from the SNe Ia discovered by the Hubble Space Telescope, which is perhaps the most accurate and reliable data, compares with that of the standard cosmology equations. The fit of Equation 1 is represented by the blue line, which lies slightly above the red line for the standard cosmology equations. The optimum $\chi^2$ for the standard cosmology equations is 18.0626, at $\Omega_m = 0.27$, $\Omega_\lambda = 0.73$, and $H_0 = 66.8$, and, for Equation 1, the optimum $\chi^2$ is 17.9686, at $\mu_p = 43.47$.

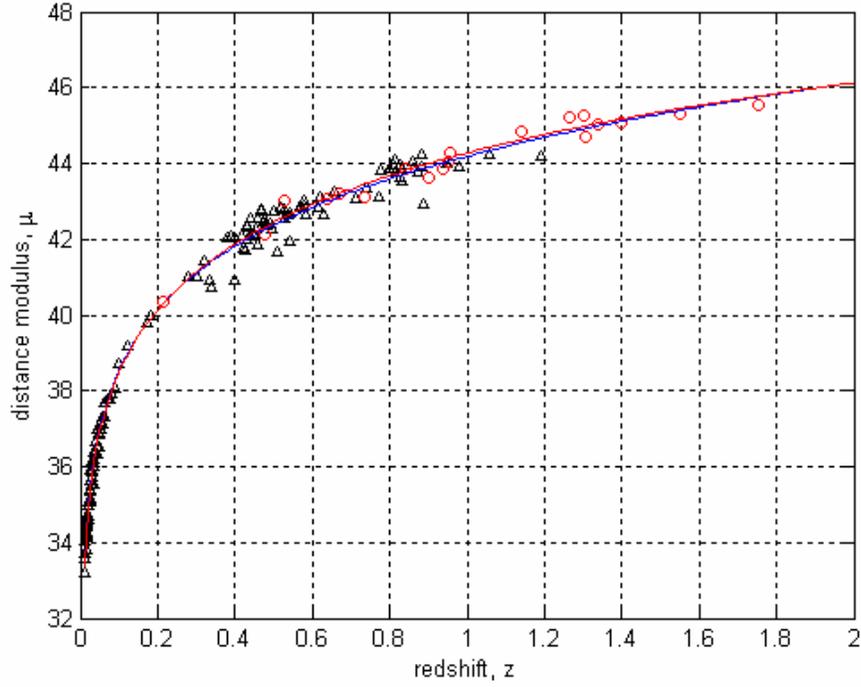

Figure 1. Distance modulus vs. redshift for Riess et al's gold dataset and the data from the HST. The triangles represent data obtained via ground-based observations, and the circles represent data obtained by the HST. The optimum fit for the standard cosmology equations is shown in red, and the optimum fit for the equation $m = m_0 + 5 \log_{10} z + 2.5 \log_{10} (z + 1)$ is shown, slightly below, in blue.

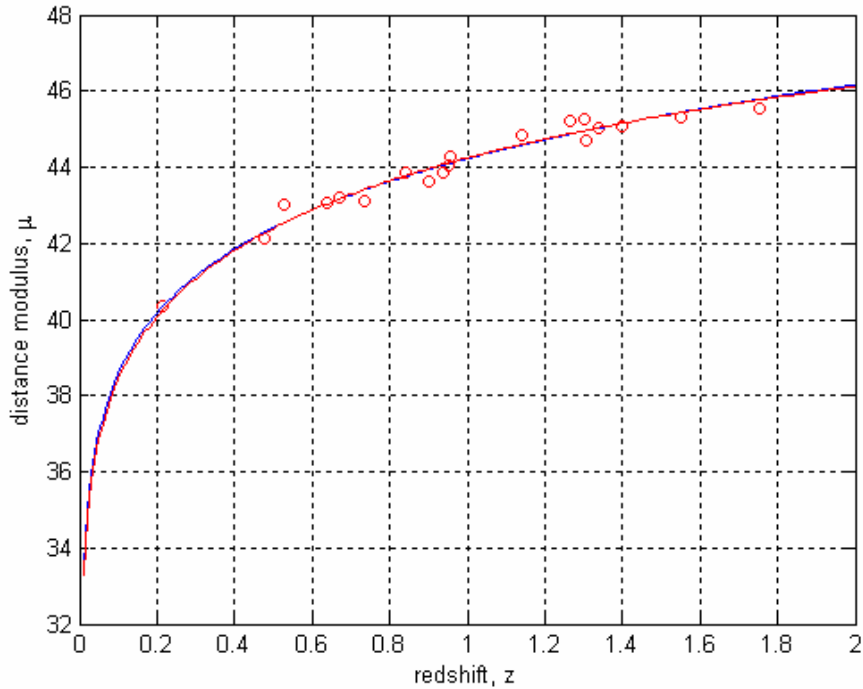

Figure 2. Distance modulus vs. redshift for Riess et al's data from the HST. The optimum fit for the standard cosmology equations is shown in red, and the optimum fit for the equation $m = m_0 + 5 \log_{10} z + 2.5 \log_{10} (z + 1)$ is shown, slightly above, in blue.

Although the standard cosmology equations fit the data in Figure 1 somewhat better than Equation 1, both are excellent fits. However, the fit of the standard cosmology equations is obtained at the expense of requiring a value of $H_0$ well below the current estimate of 71 or 72. Perhaps the best estimates for $\Omega_m$, $\Omega_\lambda$, and $H_0$ are those determined from the 2003 Wilkinson Microwave Anisotropy Probe (WMAP) satellite results[7]. These values are, $\Omega_m = 0.27$, $\Omega_\lambda = 0.73$, and $H_0 = 71$. When these values are used in the standard cosmology equations, the resulting $\chi^2$ is 296.9014, far greater than that for the corresponding fit of Equation 1, for which the resulting $\chi^2$ is 197.7256.

In Figure 2, although the $\chi^2$ values for the HST data shown are essentially equal, the $\chi^2$ value for the standard cosmology equations is, again, obtained at the expense of requiring a rather low value for $H_0$.

We note that the standard cosmology equations have two free parameters as opposed to one for Equation 1, and that could be the reason why the standard cosmology model provides a better optimum fit than Equation 1 for the data in Figure 1. When data contains an element of randomness, it is possible that the equation with more free parameters can provide a better fit, even when the equation with fewer free parameters is the one that is theoretically correct. For example, a quadratic polynomial would give a better fit than a first degree polynomial to data randomly scattered about a straight line.

### 4. Zero-energy balance in a 4-sphere

Basically, spherically closed dynamic space means conversion of spacetime with variable time and distance coordinates to dynamic space with fixed time and distance coordinates, where the line element in the fourth dimension, $ds_4 = c_4 dt$, obtains the meaning of the distance space moves in time interval $dt$ at velocity $c_4$ in the fourth dimension (in the direction of the 4-radius). The rest energy of matter gets the meaning of the energy of motion matter in space has due to the motion of space, which at the cosmological level means expansion in the direction of the radius of the 4-sphere — in a zero energy balance with the gravitation of the structure

$$M_\Sigma c_4^2 - \frac{0.776 \cdot G M_\Sigma^2}{R_4} = 0 \tag{4}$$

where $M_\Sigma$ is the total mass in space, $G$ is the gravitational constant, and $c_4$ the expansion velocity of space in the direction of the 4-radius $R_4$. The factor 0.776 in equation (2) comes from the integration of the gravitational energy in four dimensional geometry [5]. For $R_4 = 14$ billion light years and a mass density of $\rho \approx 0.55\,\rho_C$ ($\rho_C$ is the Friedman critical mass) the expansion velocity of space in the direction of the 4-radius $R_4$ obtains the value

$$c_4 = \pm\sqrt{\frac{0.776 \cdot G M_\Sigma}{R_4} = \frac{0.776 \cdot G}{R_4} 2\pi^2 R_4^3 \rho} \approx 300000 \; \left[\text{km/s}\right] \tag{5}$$

Conservation of the zero energy balance also requires that the maximum velocity in space is the velocity of space in the local fourth dimension ($c = c_4$). Assuming conservation of the primary rest energy due to the expansion of space, we find that any gravitational state and motion in space is linked to the gravitational state and motion of whole space. This means that the expansion of space (the Hubble flow) occurs not only between galaxies but also within galaxies and other local gravitational systems.

In spherically closed space, the Hubble law gets the expression

$$z = e^\alpha - 1 = \frac{D}{R_4} e^\alpha \tag{6}$$

where $\alpha$ is the angular separation of the object and the observer in spherical space (i.e. angle from the center the 4-sphere) and $D$ is the optical distance of the object (the length of the integrated path light travels in expanding 4-sphere). Equation (6) can be developed into the form

$$z = \frac{D/R_4}{1 - D/R_4} \qquad (7)$$

which relates the optical distance to the 4-radius of space. The observation angle of standard rod $r_s$ can be expressed as

$$\frac{\theta}{r_s/R_4} = \frac{z+1}{z} \qquad (8)$$

which for the angular separation of distant objects and the angular size of expanding objects obtains the Euclidean form

$$\theta = \frac{d_{(1)}}{D} = \frac{d}{(1+z)} \frac{(1+z)}{zR_4} = \frac{d/R_4}{z} = \frac{\alpha_d}{z} \qquad (9)$$

For sources like the supernova explosions, supposed to emit a fixed number of quanta in the explosion, the energy flux observed is

$$F_{obs} = \frac{F_{(e)}}{z^2(z+1)} \qquad (10)$$

which, in terms of the apparent magnitude, obtains the form of equation (1)

$$m = m_0 + 5\log_{10} z + 2.5\log_{10}(z+1) \qquad (11)$$

The Dynamic Universe model behind the equations above has been successfully tested against all major experiments on relativity including perihelion advance of planetary orbits, high precision timekeeping, and laser raging [5].

## 5. Conclusions and discussion

On the basis of our analysis, we conclude that equation (11) fits the existing SNe Ia magnitude vs. redshift data at least as well as the equation that is derived from standard cosmology theory. The derivation of equation (11) is based on very few assumptions: it assumes that 1) space is spherically closed through the fourth dimension, 2) the expansion of closed space is determined by a zero energy balance of motion and gravitation, and 3) the energy mass has due to the motion of space in the fourth dimension is conserved in all interactions in space. Once these assumptions are accepted, the mass density of space is determined by the zero energy balance and the apparent magnitude can be expressed as a function of the redshift only. The fit of equation (11) also means that no dark energy is needed to explain the supernova observations.

The interpretation of the cosmic expansion as a zero energy process of spherically closed three dimensional space can be regarded as a four dimensional solution of the Einstein-deSitter expansion. The expansion continues forever and the velocity of the expansion in the fourth dimension decreases until it becomes zero at infinity. Since the rest energy of mass is bound to the velocity of space in the fourth dimension, also the rest energy of matter goes to zero at infinity. The zero energy process is symmetric like a spherical pendulum in the fourth dimension; the energy of motion is gained against the release of gravitational energy in a contraction phase and it is released back to gravitational energy in an expansion phase following the culmination of the contraction in a singularity.

In the coordinate system based on a dynamic 4-sphere the fourth dimension is geometrical in its nature, allowing a universal time coordinate. As a consequence of the third assumption above, the conservation of energy in interactions in space, the frequencies of atomic emission and absorption become linked to local motion and gravitation, which means that the concept of proper time is replaced by "proper frequencies". In addition, the velocity of light becomes linked to the local gravitation, which means that, although not a constant in dynamic space, the velocity of light is *observed* as being constant in local measurements.